# Ultrafast Liquid Water Transport through Graphene-based Nanochannels Measured by Isotope Labelling

Pengzhan Sun[1], He Liu[2], Kunlin Wang[1], Minlin Zhong[1], Dehai Wu[2], Hongwei Zhu[1,3]

[1]School of Materials Science and Engineering, State Key Laboratory of New Ceramics and Fine Processing, Tsinghua University, Beijing 100084, China

[2]Department of Mechanical Engineering, Tsinghua University, Beijing 100084, China

[3]Center for Nano and Micro Mechanics, Tsinghua University, Beijing 100084, China

## ABSTRACT

Graphene-based laminates, with ultralong and tortuous nanocapillaries formed by simply stacking graphene flakes together, have great promises in filtration and separation. However, the information on liquid water trans-membrane permeation is lacking, which is the most fundamental problem and of crucial importance in solution-based mass transport. Here, based on isotope labelling, we investigate the liquid water transportation through graphene-based nanocapillaries under no external hydrostatic pressures. Liquid water can afford an unimpeded permeation through graphene-based nanochannels with a diffusion coefficient 4~5 orders of magnitude larger than through sub-micrometer-sized polymeric channels. When dissolving ions in sources, the diffusion coefficient of ions through graphene channels lies in the same order of magnitude as water, while the ion diffusion is slightly faster than water, indicating that the ions are mainly transported by fast water flows and the delicate interactions between ions and nanocapillary walls also take effect in the accelerated ion transportation.

Mass transport through materials with nanoscaled pores or channels is currently an attractive subject with significant implications in nanofluidic device design and the fundamental understanding of fluids at nanoscale.[1,2] Among them, carbon nanomaterials, *e.g.* carbon nanotubes[3-7] and graphene oxide (GO) membranes[8,9], with numerous hydrophobic graphitic nanochannels inside, can afford an almost frictionless transport of water molecules. More recently, GO membranes are reported with the ability to separate different ions in solutions with an ultrafast speed based on the physical size effect of nanocapillaries[10,11] and the diverse interactions between ions and GO membranes[12-15]. In combination with the characteristics of easy to synthesize and scale up,[16,17] GO membranes are believed to have great promises in large-scale industrial filtration and separation. Typically, in contrast to GO layers with a nanometer-sized thickness,[18-21] in which mass transport happens mostly through the structural defects within GO flakes, as-synthesized micrometer-thick GO membranes are impermeable to most of the liquids and gases (including helium). Exceptionally, water vapors can afford an unimpeded permeation through the $sp^2$ nanocapillary networks formed by connecting all the stacking layers across the whole laminates,[8,9] and small ions (hydrated radii < 4.5 angstroms) can also permeate through with an ultrafast speed.[10] In spite of these exciting results obtained with GO membranes, the permeation behavior of liquid water, which is a much more complex system than the gaseous state, is unclear; that is the most fundamental problem and of crucial importance for mass transport in solutions.

The investigation of liquid water permeation under no external hydrostatic pressure remains to be a challenge, because no detectable differences in both macroscopic and microscopic levels are present between the sources and drains separated by the central GO membranes. To overcome this issue, certain amount of deuterium oxide ($D_2O$) is dissolved as a tracer to label the source water and the trans-membrane permeation of $D_2O$ is investigated to extrapolate that of water. Based on this isotope labelling technique, the permeation behavior of liquid water through the nanocapillaries within GO membranes prepared by vacuum-filtration of GO colloidal solutions is studied in this work.

## RESUTLS AND DISCUSISON

Initially, GO sheets were synthesized by the modified Hummers' method starting from natural graphitic flakes, which were exposed to potassium permanganate, sodium nitrite and concentrated sulfuric acid subsequently.[22] Figure S1 shows the SEM and AFM characterizations of as-synthesized GO flakes. It reveals that the GO flakes are single-layered with a typical lateral dimension of ~1 μm. After dissolving the GO flakes in water by sonication, micrometer-thick GO membranes were prepared by vacuum filtration with the commonly used cellulose microfilters (pore size: ~220 nm, porosity: ~80%), as illustrated in Fig. 1a. Typically, a network of ultralong and tortuous $sp^2$ nanocapillaries is believed to form within the micrometer-thick GO laminates, which is responsible for the transport of water molecules and ions.[8-11,23] This has been further confirmed by varying the lateral dimensions of GO flakes within the membranes from micrometer to nanometer, which yields an enhancement of ion transportation.[13,15] In contrast, for the case of nanometer-thick GO membranes, in which only few layers of GO flakes are stacked together, the expected continuous nanocapillary networks are hardly to form and the mass transport is dominated by the structural defects within the GO flakes.[18,19] Hence, for the micrometer-thick GO membranes used in all the experiments here, we believe that the $sp^2$ nanocapillaries formed within the laminates are mainly responsible for the water and ion transportations, as sketched in Fig. 1a. After the membrane preparation procedure, the GO membrane with the cellulose microfilter underneath (named as "GOCM") was assembled into a home-made permeation apparatus, as shown in Fig. 1b.[24] Note that the GO membranes were not detached off from the microfilters because the underneath polymeric substrates could provide a valuable mechanical strength enhancement to ensure that the GO membranes were not cracked during the water permeation experiments. The intactness and continuity of the GO membranes were also checked by optical microscopy before and after experiments. To extract the effect of the substrates on water permeation, control experiments were conducted with blank microfilters, as shown in Fig. 1b. In the water permeation experiments, 100 mL of $D_2O$ labelled water with various mass concentrations and deionized water were injected into the source and drain reservoirs, respectively. The whole water permeation process was conducted under mild magnetic stirring, as shown in Fig. 1c, to avoid possible $D_2O$ concentration gradients around the membranes. The filtrates in drains were examined by Fourier transform infrared (FTIR) spectroscopy to afford accurate concentrations of $D_2O$ tracers (discuss later), based on which the permeation behavior of liquid water could be extrapolated. Figure 1d exhibits the interface topography between GOCM and microfilter obtained by white light interference microscope, from which the thickness of the GO membranes prepared by vacuum-filtrating 25 mL, 0.1 mg/mL GO solutions can be determined as ~4 μm, which has also been confirmed by stylus profilometry, as shown in Fig. S2. The thickness of the cellulose microfilters used here was evaluated by the cross-section optical microscopy, as shown in Fig. 1e, which reveals that the microfilters possess a thickness of ~115 μm.

In terms of measuring the concentration of $D_2O$ tracers in drain solutions, FTIR spectroscopy was utilized. Figure 2a shows an example of the FTIR spectrum of 50 wt% $D_2O$ solution. Typically, gaseous water is a nonlinear three-atomic molecule and its FTIR spectrum exhibits three characteristic peaks located at $v_1$ = ~3652 $cm^{-1}$, $v_2$ = ~1596 $cm^{-1}$ and $v_3$ = ~3756 $cm^{-1}$, respectively. However, due to the strong hydrogen bonding effect in liquid water, the extension vibration modes of $v_1$ and $v_3$ overlap together to form a wide peak located at ~3440 $cm^{-1}$, while the angular vibration mode $v_2$ is located at ~1645 $cm^{-1}$. On the other hand, in the case of $D_2O$, due to the slightly larger atomic mass of deuterium than hydrogen, the overlapped peak corresponding to extension vibration modes ($v_1$' and $v_3$') shifts to ~2540 $cm^{-1}$ and the angular vibration mode $v_2$' shifts to ~1210 $cm^{-1}$ (Fig. 2a). Therefore, based on the absorption intensity of the characteristic peak located at ~2540 $cm^{-1}$ in the FTIR spectra, the accurate concentration of $D_2O$ tracers in drains can be determined. Firstly, the FTIR spectra of $D_2O$ aqueous solutions with fixed mass concentrations were carried out to quantify the function between the absorption intensity at ~2540 $cm^{-1}$ and the mass concentration of $D_2O$ in solutions (Fig. S3). The obtained FTIR spectra were baselined and normalized with the overlapped peak located at ~3440 $cm^{-1}$ (Fig. 2a and Fig. S3) and the absorption intensities located at ~2540 $cm^{-1}$ were plotted as a function of the mass concentrations, as shown in Fig. 2b. It reveals that the absorption intensity at ~2540 $cm^{-1}$ ($A$) varies linearly with the mass concentration of $D_2O$ ($c_m$), following a function of $A = 0.01983\ c_m$. With this linear function, the $D_2O$ labelled water permeation properties through GO membranes can be investigated and the relationships of $D_2O$ mass transport *versus* time are plotted in Figs. 2c-f. For each group of experiments, where water sources with different $D_2O$ concentrations were allowed to permeate through GOCM and blank microfilter respectively, at least three runs were repeated. Excellent reproducibility could be obtained; the data can be well fitted into a linear relationship and the error bars are within the size of the data balls. As shown in Figs. 2c-f, substantial deviations occur between the $D_2O$ permeations through GOCM and microfilter. With the gradual decrease of the $D_2O$ source concentrations, the differences in $D_2O$ permeations through GOCM and its control membrane become smaller. Surprisingly, when the source concentration of $D_2O$ is down to 10 wt%, the permeations *versus* time of trace quantities of $D_2O$ through these two membranes nearly coincide together, just like in the absence of GO membrane; that is, for the blank microfilter. This indicates that liquid water can afford an almost unimpeded permeation through GO membranes, just the same as water vapor,[8] and also in consistent with the mass transport properties observed in carbon nanotubes.[3-7] Additional experiments with even lower $D_2O$ concentrations were also performed and we found that the amount of $D_2O$ in drains was failed to detect accurately, indicating that 10 wt% is a lower bound for the observation of $D_2O$ labelled water flow through GO membranes.

As shown in Figs. 2c-f, after 8 h of permeation, the amount of $D_2O$ in drains only increases to < 5% compared to source concentrations. This means that water trans-membrane permeation can be treated as a quasi-static process, in which the Fick's first law can be utilized to calculate the diffusion coefficients of $D_2O$ through GO membranes and microfilters, as illustrated in Fig. 3a.[24] Firstly, the flow rates of $D_2O$ through GOCM and blank microfilter membranes were calculated, as shown in Fig. 3b. It reveals that reducing the source concentration yields the decrease of the $D_2O$ permeation rates as well as the deviations between the permeations through GOCM and microfilter membranes. Notably, the calculated water fluxes here are all significantly greater (1~2 orders of magnitude) than the results obtained by Joshi, et al. (~0.2 L $m^{-2}$ $h^{-1}$).[10] In their work, the water flow rates were calculated based on the changes in the liquid level differences during the permeation of 1 M sucrose,[10] which corresponds to a osmosis pressure of ~25 bar at room temperature ($\pi = cRT$, where $\pi$ is the osmosis pressure, $c$ is the source concentration, $R$ is the gas constant and $T$ is the temperature, assuming the van't Hoff factor is 1). In the case here, the transport of $D_2O$ is driven by much larger osmosis pressures (≈125~930 bar, corresponding to the $D_2O$ concentration of 10~70 wt%), which is in exact agreement with the work by Joshi, et al.[10] With the $D_2O$ flow rates through GOCM and microfilter membranes (Fig. 3b), the diffusion coefficients of $D_2O$ through microfilter, GO and GOCM membranes can be calculated, as plotted in Fig. 3c.[24] It reveals that the diffusion coefficients of $D_2O$ through GO membranes are smaller than through microfilters by ~1 order of magnitude when the source concentration is varied from 10 wt% to 70 wt%. At first glance, one may conclude that such a small diffusion coefficient of water through GO membranes makes it even less promising than the commonly used several hundred micrometer-thick cellulose microfilters. However, if the microstructure of GO membranes is considered, we can draw a rather different conclusion. In view of the structure of GO sheet, nanosized $sp^2$ clusters are distributed randomly within the $sp^3$ C-O matrix.[25,26] When a large amount of GO flakes are stacked together to form the micrometer-thick laminate, numerous millimeter-long graphitic nanocapillaries can be formed by connecting the $sp^2$ clusters across all the stacking layers,[24] which is responsible for the transport of water, as illustrated in the inset of Fig. 3d. Therefore, we propose that liquid water diffuses mostly through the nanocapillaries formed within the GO laminates and the permeation through the structural defects is neglected because of mutual stacking, just in consistent with the previous work.[8,9,10,17,18,23] Similarly, in the case of microfilters with porosities of ~80%, we propose that water diffusions happen mostly through the sub-micrometer-sized pores (~0.22 μm) within the matrix. Based on the microstructures of GO and cellulose microfilters, the diffusion coefficients of $D_2O$ through the channels within GO and microfilters can be calculated, as plotted in Fig. 3d.[24] Surprisingly, the water diffusion coefficients through the nanochannels within GO membranes are 4~5 orders of magnitude greater than through the sub-micrometer-sized cellulose pores, which is in agreement with the concept of frictionless transport of water through the inner graphitic channels within carbon nanotubes.[3-6] Note that the calculated water diffusion coefficients through the nanochannels within GO membranes are lower bounds because recent simulation results have predicted that the functionalized C-O regions within the GO sheets suppress water permeation.[27,28]

Specially, the water permeation properties through GO membranes when dissolving ions in solutions were investigated based on 0.1 M $MgCl_2$ sources and 30 wt% $D_2O$ tracers, as shown in Fig. 4. Firstly, the 0.1 M $MgCl_2$ source solution was labelled with $D_2O$ tracers to study the forward water transport in the presence of ions, as illustrated in Fig. 4a. It reveals that the water permeation is slightly reduced when ions are dissolved in source solution (Figs. 4b and c). Previous first-principle calculations have demonstrated that the formation of ice bilayer within the interlayer spacing and its melt transition at the edges of the flakes are responsible for the rapid water transport through GO membranes.[9] Herein, in the presence of ions in nanocapillaries, we propose that the existent ions cause significant distortion to the ordered ice bilayer, further yielding the reduced water permeation rates, as sketched in the inset of Fig. 4b. Next, the drain solution was labelled with 30 wt% $D_2O$ to study the water permeation from drain to source in the presence of source ions, as illustrated in Fig. 4d. Notably, it reveals that the water permeation from drain to source is slightly faster than from source to drain when 0.1 M $MgCl_2$ is present in source solutions (Figs. 4e,f). This can be attributed to the semipermeable effect of GO membranes,[10] through which water are tended to be transported against the concentration gradients. On the other hand, the ion trans-membrane permeations in the presence of $D_2O$ tracers are shown in Figs. S4a-d, indicating that the $D_2O$ molecules in sources or drains have neglected effect on ion permeations. The calculated ion diffusion coefficients for the entire membranes and the channels within the membranes are found to be in the same order of magnitude as water, as shown in Figs. S4e-h. These results suggest that the ions in sources are mainly transported by the fast water flows through GO membranes. Notably, the diffusion coefficients of ions through graphene-based nanochannels are slightly larger than water, indicating that the delicate interactions between ions and GO membranes[12-15] also take effect in the accelerated ion transportation.

## CONCLUSION

In summary, liquid water transport through graphene-based nanochannels without external hydrostatic pressures has been investigated, which remains to be a challenge and most fundamental problem for solution-based mass transport. Based on isotope labelling, we have shown that liquid water can undergo an almost frictionless permeation through the millimeter-long nanocapillaries in GO membranes, just similar to the case of carbon nanotubes and water vapors in GO laminates. However, the study on liquid water permeation makes more sense than the gaseous case because it can throw light upon the mechanism of ultrafast ion transportation through GO membranes, which is of crucial importance for mass transport in solutions. Also the solution-processed GO membranes are readily to fabricate and scale up, which are more promising than the CVD-synthesized carbon nanotubes. The results present here may indeed lay the foundation on nanofluidic device design and fast mass transportation based on engineering the nanochannels within GO membranes.

## MATERIALS AND METHODS

**Preparation of GO membranes.** Monolayer GO flakes were synthesized by the modified Hummers' methods from natural graphite, which was treated with potassium permanganate, sodium nitrite and concentrated sulfuric acid subsequently.[22] Figure S1 exhibits the SEM and AFM images of the as-synthesized GO sheets, which reveals that the GO flakes are single-layered with a typical lateral dimension of ~1 μm. As-synthesized GO flakes were re-dispersed in deionized water by sonication to form the 0.1 mg/mL GO preparation solutions. After that, the micrometer-thick GO membranes were fabricated by vacuum-filtrating 25 mL (0.1 mg/mL) GO colloidal solutions using the commercial cellulose microfilters. The polymeric microfilters possess pores with diameters of ~0.22 μm and a porosity of ~80%. After the vacuum-filtration process, the GO membranes with the microfilters underneath (named as "GOCM") were dried in air at 55$^{o}$C for 24 h before use. This drying procedure is important because the GO membranes without drying are poorly adhered to the microfilters, further leading to the ease of exfoliation of GO membranes during the penetration tests in which the GOCM samples were immersed in water for at least 8 h. As a control experiment, the blank cellulose microfilters were also dried in air at 55$^{o}$C for 24 h before use.

**Permeation test setup.** The isotope labelled water permeation tests were done with a self-made penetration apparatus, as shown in Figures 1b and c. Briefly the source reservoir and drain reservoir were separated by a plastic plate with an aperture (~5 mm in diameter) in the center. A piece of GOCM (or blank microfilter) was sealed with double-sided copper tape onto the aperture within the plastic plate. The double-sided copper tape possessed the same sized hole (~5 mm in diameter) in the center so that the GOCM (or blank microfilter) could directly connect the solutions in source and drain reservoirs with an effective diffusion area of ~19.6 mm$^2$. The regions around the effective diffusion membranes were also sealed with copper tapes as a protection, as shown in

Figure 1b. Before and after the penetration experiments, the intactness and continuity of GO membranes were checked by optical microscopy to ensure that no cracks were formed on the GO membranes and the results obtained by isotope labelled water permeation reflected the true behavior of GO membranes.

During the water permeation experiments, 100 mL of $D_2O$ labelled water with various $D_2O$ mass concentrations and deionized water were injected into the source and drain reservoirs with the same speed. Mild magnetic stirrings were applied to both the source and drain solutions during the whole permeation process to avoid possible $D_2O$ concentration gradients near the membranes. During the permeation process, equivalent amount of solutions in sources and drains were taken out at regular intervals (typically 2 h) for characterization to avoid the external hydrostatic pressures across the membranes caused by the unequal liquid levels in sources and drains. The filtrates in drains were then examined by Fourier transform infrared (FTIR) spectroscopy to afford accurate concentrations of $D_2O$ tracers, based on which the permeation behavior of water through GO membranes could be studied.

**Characterizations.** As-synthesized GO flakes were characterized by scanning electron microscope (SEM, LEO 1530, 10kV) and atomic force microscopy (AFM, Agilent 5100). The thickness of as-prepared GO membranes was determined by white light interferenc microscope (MicroXAM-1200) and stylus profilometry (Ambios XP-1). The thickness of cellulose microfilters was evaluated by optical microscope (ZEISS, Axio Scope.A1). The accurate concentrations of $D_2O$ were measured by Fourier transform infrared (FTIR) spectroscopy (Nicolet 6700FTIR). In terms of ions dissolving in source solutions, the concentrations of $Mg^{2+}$ cations were measured by atomic emission spectroscopy (IRIS Intrepid II).

**Figures**

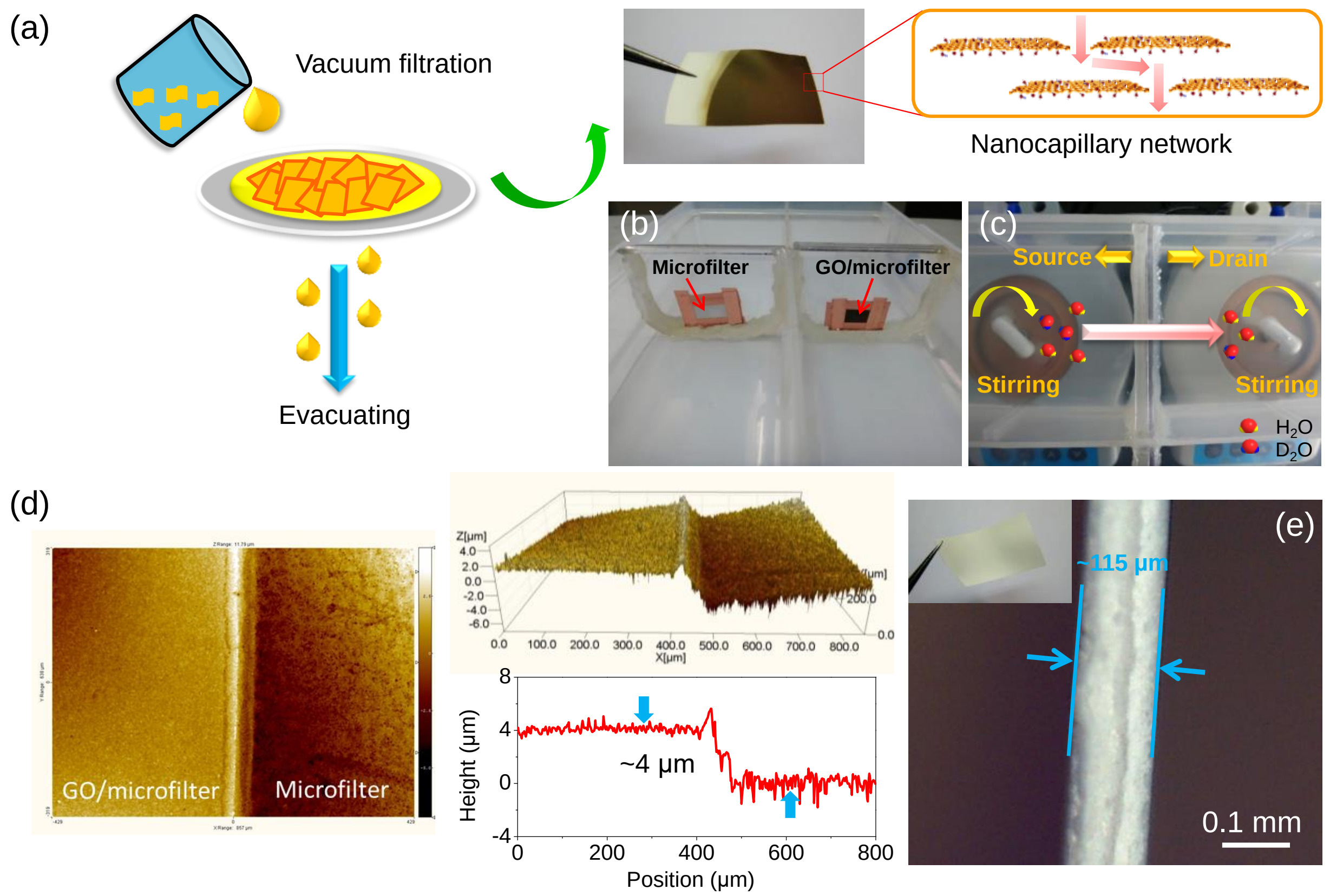

**Fig. 1. Experimental setup and membrane characterizations.** (**a**) Left panel: schematic drawing for the fabrication of GO membranes by vacuum-filtration. Right panel: a photograph for the as-synthesized GO membrane and a schematic diagram for its cross-sectional structure. (**b** and **c**) Photographs for the home-made permeation apparatus and the $D_2O$ labelled water trans-membrane permeation process. (**d**) White light interferenc characterizations for the interface between GOCM and microfilter membrane. (**e**) An optical image for the cross-section of cellulose microfilter. The inset shows a photograph of the microfilter used in the experiments.

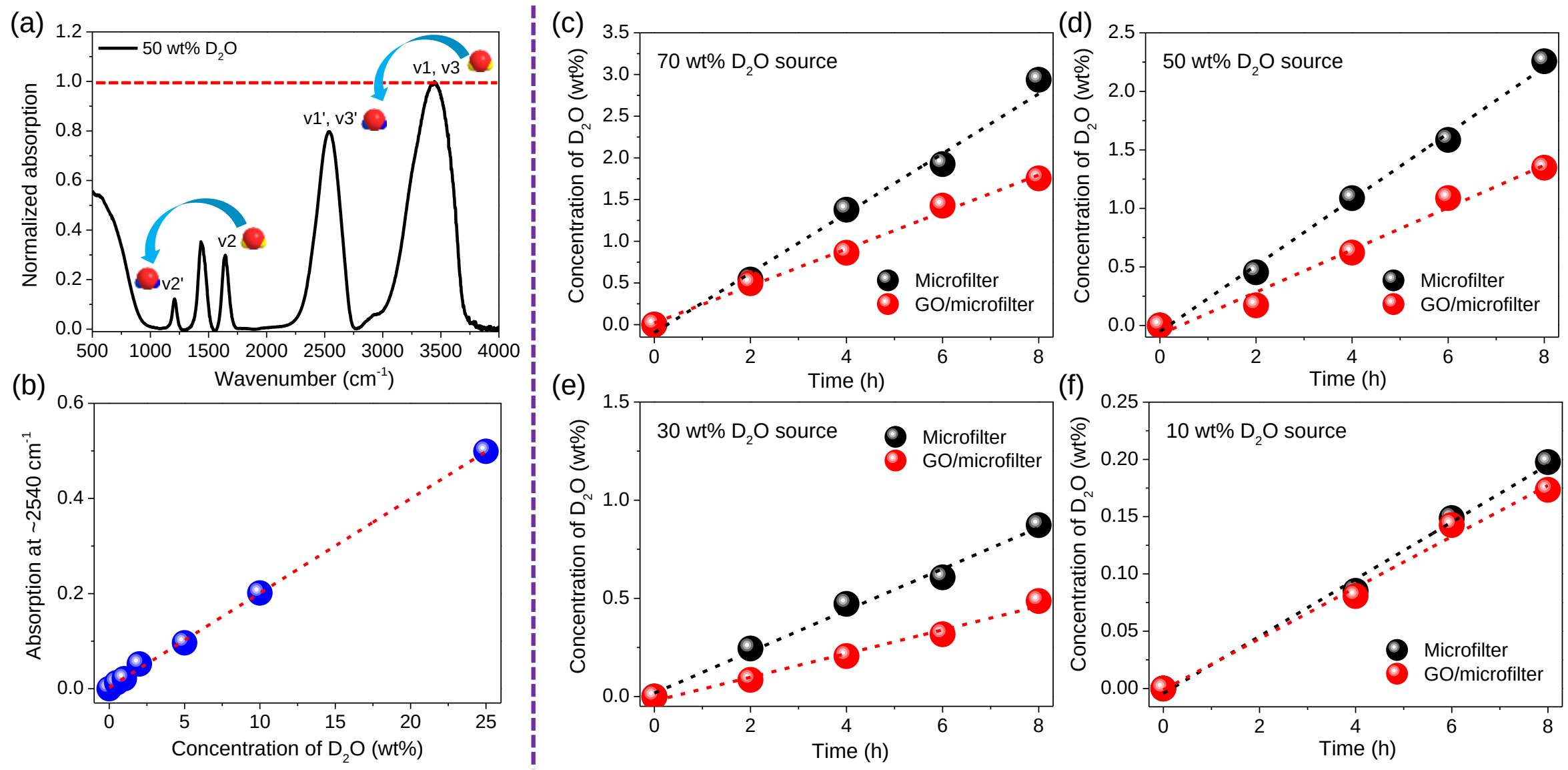

**Fig. 2. $D_2O$ labelled water trans-membrane permeation measured by FTIR spectra.** (a) FTIR spectrum of a 50 wt% $D_2O$ solution to characterize the peaks assigned to $H_2O$ and $D_2O$. (b) Functional relationship between the absorption intensity at ~2540 $cm^{-1}$ and the mass concentration of $D_2O$ in solutions. $D_2O$ labelled water trans-membrane permeation through GOCM and microfilter membranes with various $D_2O$ source concentrations: (c) 70 wt%, (d) 50 wt%, (e) 30 wt% and (f) 10 wt%. All the experiments were repeated for at least three times. The variations of the data are within the size of the data balls.

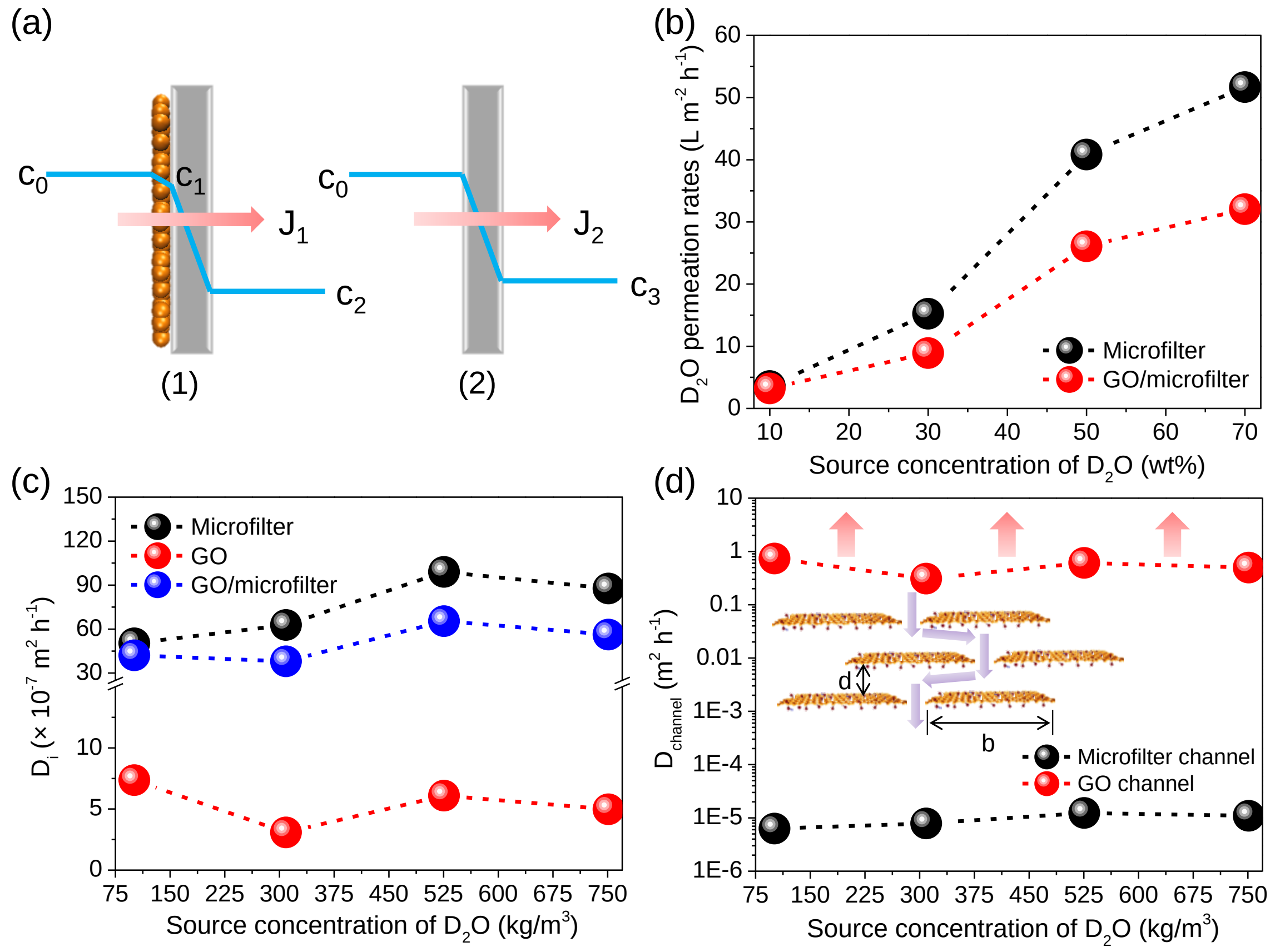

**Fig. 3. Water permeation rates and diffusion coefficients.** (a) Schematic diagram for the calculation of diffusion coefficients through GO and microfilters by Fick's first law. (b) $D_2O$ permeation rates through GOCM and blank microfilter membranes as a function of the source concentration. (c) The calculated $D_2O$ diffusion coefficient through entire GO, microfilter and GOCM membranes as a function of the $D_2O$ source concentration. (d) $D_2O$ diffusion coefficient through the channels within GO and microfilters as a function of the $D_2O$ source concentration. The inset shows a schematic diagram for the cross-section of GO membranes used for calculation. The arrows indicate the lower bounds for the water diffusion coefficients through the nanocapillaries within the GO membranes.

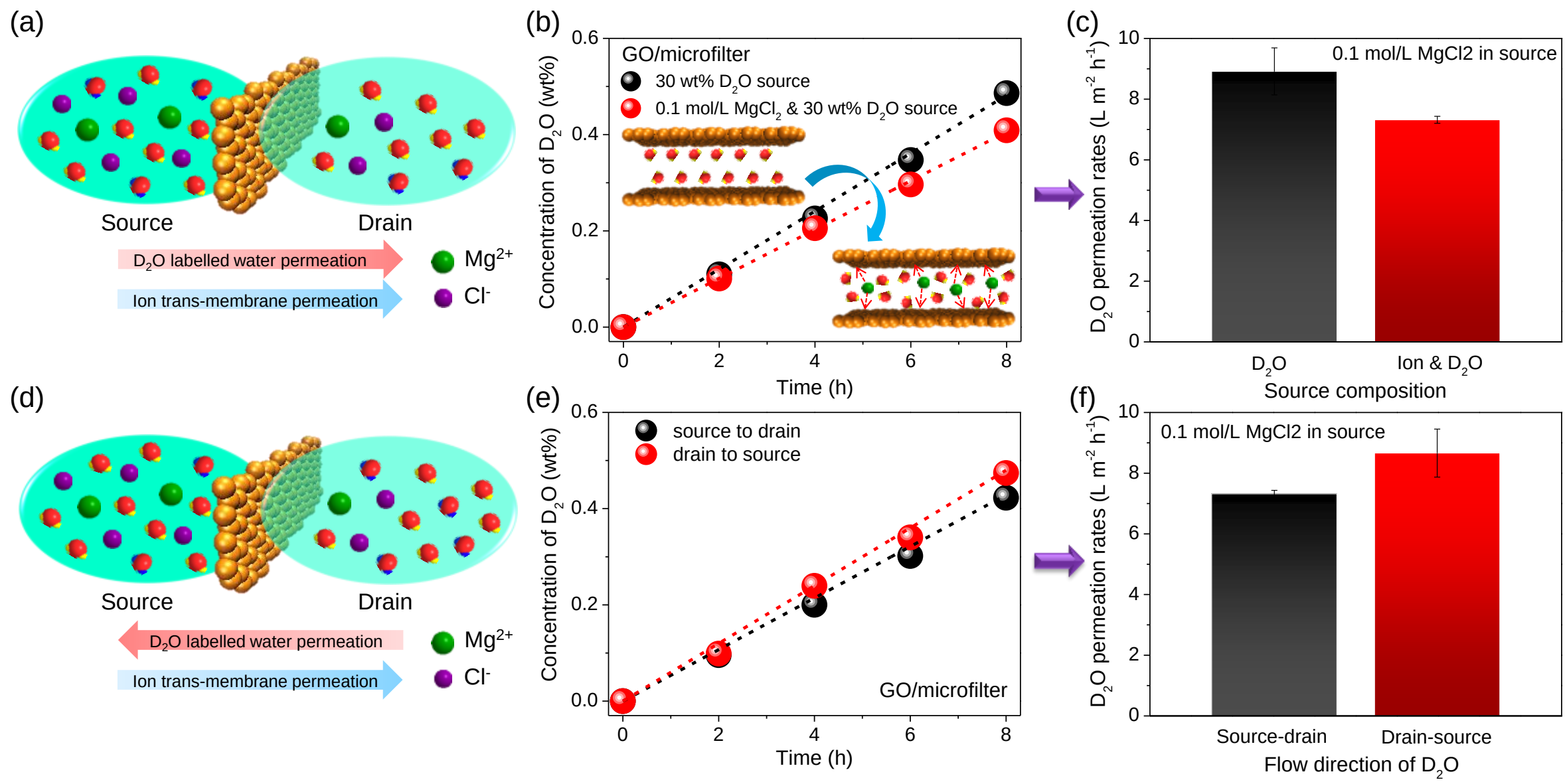

**Fig. 4. $D_2O$ labelled water permeation in the presence of 0.1 M $MgCl_2$ in sources.** (a) Schematic drawing for the labelling of 0.1 M $MgCl_2$ source solution by 30 wt% $D_2O$ tracers. (b) Water permeations through GOCM membranes with and without ions in sources. The inset shows the possible mechanism for water permeation in the presence of ions. (c) Water permeation rates through GOCM membranes with and without ions in sources. (d) Schematic diagram for the labelling of drain solutions by 30 wt% $D_2O$ tracers when dissolving 0.1 M $MgCl_2$ in source solutions. (e) Water permeations through GOCM membranes in both directions when dissolving ions in sources. (f) Water permeation rates through GOCM membranes in both directions in the presence of source ions.

Supplementary Information

# Ultrafast Liquid Water Transport through Graphene-based Nanochannels Measured by Isotope Labelling

Pengzhan Sun[1], He Liu[2], Kunlin Wang[1], Minlin Zhong[1], Dehai Wu[2], Hongwei Zhu[1,3]

[1]School of Materials Science and Engineering, State Key Laboratory of New Ceramics and Fine Processing, Tsinghua University, Beijing 100084, China

[2]Department of Mechanical Engineering, Tsinghua University, Beijing 100084, China

[3]Center for Nano and Micro Mechanics, Tsinghua University, Beijing 100084, China

**This file includes:**

SOM text

Figs. S1 to S4

References

**SOM text**

## *1. The water diffusion coefficients through GO, microfilter and GOCM membranes*

The diffusion coefficients of $D_2O$ through GO, GOCM and blank microfilter membranes were calculated by Fick's first law based on the fact that the increase of the amount of $D_2O$ in drains was less than 5% compared to the source concentration after such a long 8 h of permeation, as shown in Figures 2c-f.

As illustrated in Figure 3a, the flux of $D_2O$ through GOCM (1) can be expressed as the following equation:

$$J_1 = D_G \frac{c_0 - c_1}{l_G} = D_M \frac{c_1 - c_2}{l_M} \approx D_M \frac{c_1}{l_M} = D_{total} \frac{c_0 - c_2}{l_G + l_M} \approx D_{total} \frac{c_0}{l_G + l_M} \qquad (1)$$

where $J_1$ is the flux of $D_2O$ through GOCM membrane, $D$ is the diffusion coefficient, $c_0$, $c_1$ and $c_2$ are concentrations of $D_2O$ in source, at the interface between GO and microfilter and in drain, respectively. $l$ is the thickness of the membranes. The subscripts "*G*", "*M*" and "*total*" represent GO, microfilter and GOCM, respectively. The value of $c_2$ is much smaller than those of $c_0$ and $c_1$, so it can be neglected. On the other hand, the flux of $D_2O$ through microfilter (2) can be expressed as follows:

$$J_2 = D_M \frac{c_0 - c_3}{l_M} \approx D_M \frac{c_0}{l_M} \qquad (2)$$

where $J_2$ is the flux of $D_2O$ through microfilter and $c_3$ is concentration of $D_2O$ in drain in the case of water permeation through microfilter. The value of $c_3$ is much smaller than that of $c_0$, so it can be neglected. Combining Eqs. (1) and (2), one can calculate the diffusion coefficients of $D_2O$ through GO, microfilter and GOCM membranes, respectively, as shown in Figure 3c.

## 2. *Evaluation of the water diffusion coefficients through GO and microfilter channels*

Considering the microstructures of GO membranes and microfilters, the water diffusion coefficients through the channels within GO membranes and microfilters can be evaluated according to previous methods.[1,2] As discussed in the main text, a graphitic nanocapillary network can be formed by connecting the $sp^2$ clusters across all the stacking layers together, which is mainly responsible for the transport of water molecules, as illustrated in the inset of Figure 3d. The lateral dimensions of GO flakes ($b$) within the membranes can be evaluated as ~1 μm according to the SEM and AFM characterizations in Figure S1, while the interlayer distances between GO flakes ($d$) can be evaluated as ~1 nm according to the previous XRD analyses.[2-6] The thickness of our GO membrane samples ($l$) is ~4 μm, as shown in Figure 1d and S2. Therefore, the effective length of graphene nanochannels can be evaluated as $l_{\mathrm{eff}} = bl/d$ = 4 mm, while the effective diffusion area can be evaluated as $A_{\mathrm{eff}} = Adb/b^2 = Ad/b = 0.1\%A$, which occupies only a tiny fraction of the total GO membrane area. Based on these parameters, the water diffusion coefficient through the nanochannels within GO membranes can be calculated based on the following equation:

$$D\frac{A}{l} = D_{channel}\frac{A_{eff}}{l_{eff}} \tag{3}$$

According to Eq. (3), $D_{channel}$ can be calculated as $10^6\ D_G$. Note that the evaluated $D_{channel}$ for GO membranes here is a lower bound because the functionalized C-O regions suppress the through-permeation of water seriously, according to the recent simulation studies.[7,8]

Similarly, for the case of cellulose microfilters used here, the pore size is ~0.22 μm and the porosity is ~80%. The effective diffusion area $A_{\mathrm{eff}}$ can be calculated as 80%$A$ while the effective length of diffusion channels $l_{\text{eff}}$ is equivalent to the thickness of microfilters (~115 μm, as shown in Figure 1e). According to Eq. (3), the water diffusion coefficient through microfilter channels can be calculated as 1.25 $D_M$, which is comparable to bulk case. The results for water diffusion coefficients through the channels within GO membranes and microfilters are plotted in Fig. S3d, revealing that the diffusion coefficients of water through graphene-based nanochannels are 4~5 orders of magnitude greater than the bulk diffusion case.

## Supplementary figures

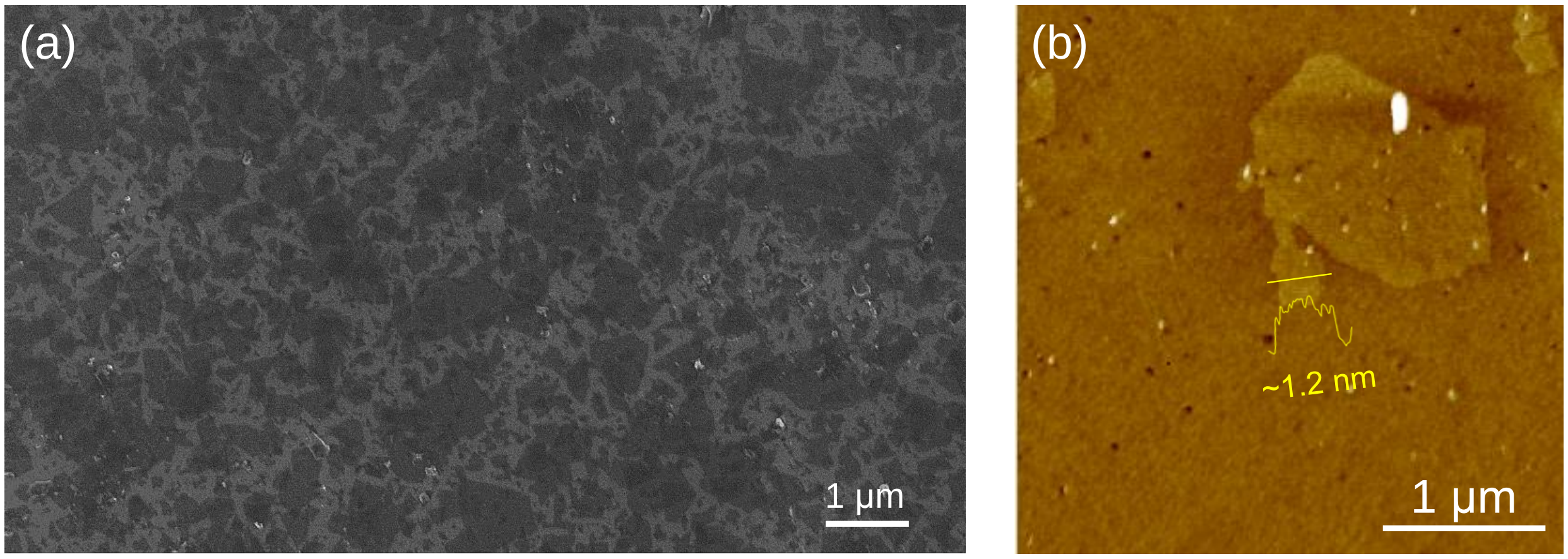

**Fig. S1.** SEM (a) and AFM (b) characterizations of as-synthesized GO flakes. The inset in (b) shows the height profile for the corresponding yellow line.

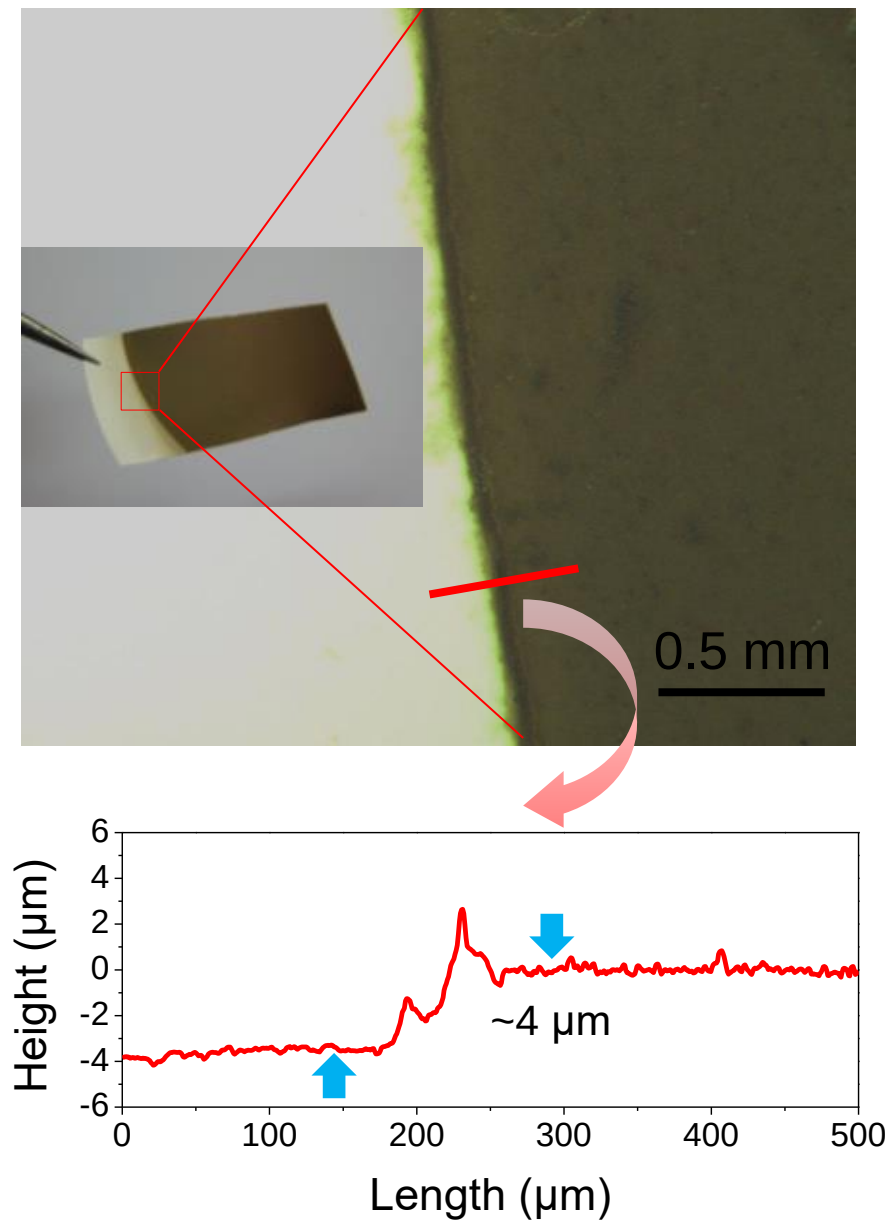

**Fig. S2. Measuring the thickness of the GO membrane by stylus profilometry.** The photograph inset is an example of the as-fabricated GO membrane with the cellulose microfilter underneath. The red line in the optical image shows the measured region.

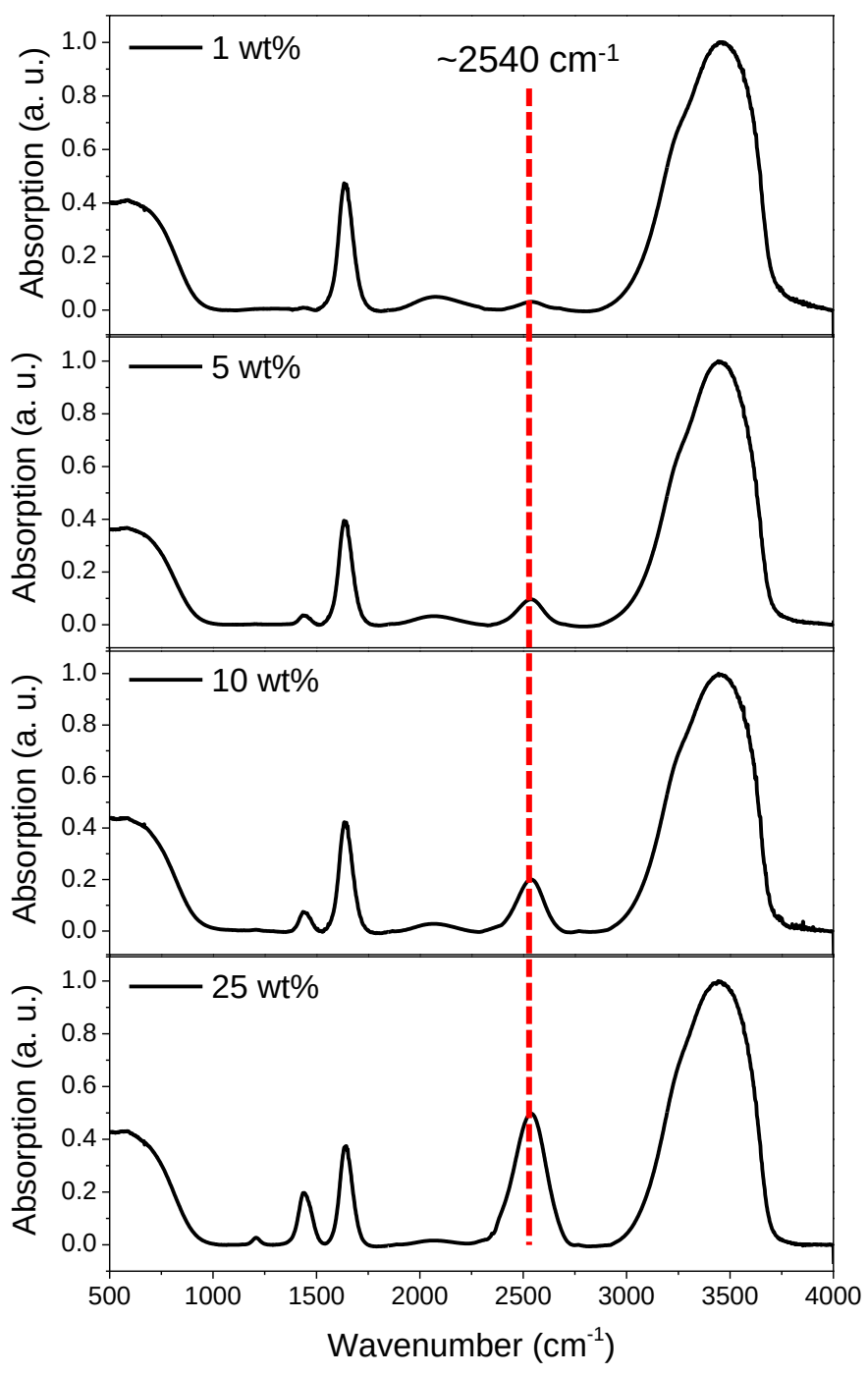

**Fig. S3. FTIR spectra of the $D_2O$ solutions with fixed mass concentrations**: 1 wt%, 5 wt%, 10 wt% and 25 wt%, respectively.

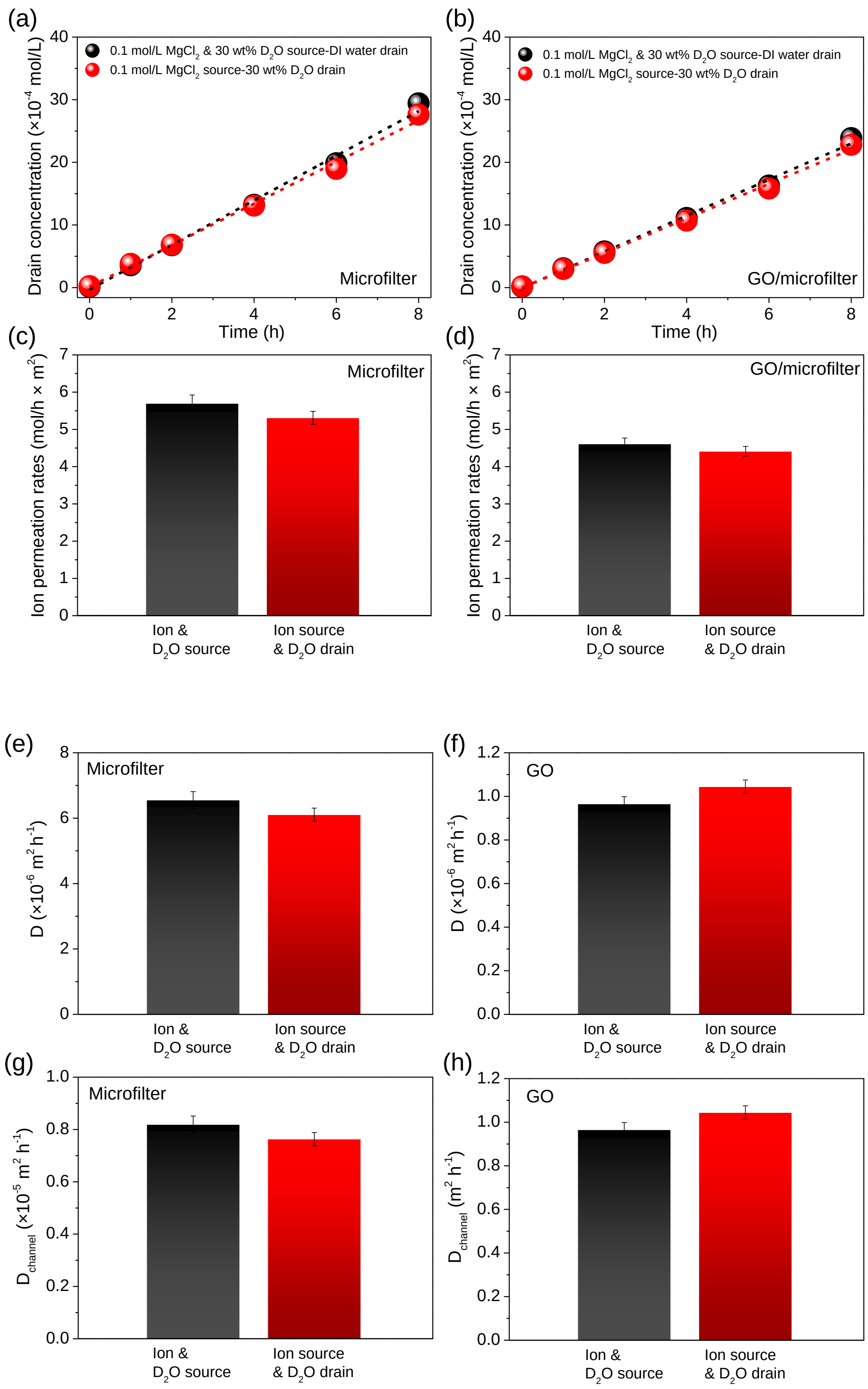

**Fig. S4. Ion permeation through microfilter and GOCM membranes.** $Mg^{2+}$ ion permeations through microfilters (a) and GOCMs (b) in the presence of 30 wt% $D_2O$ in sources and drains,

respectively. $Mg^{2+}$ ion permeation rates through microfilters (c) and GOCMs (d) in the presence of 30 wt% $D_2O$ in sources and drains, respectively. $Mg^{2+}$ ion diffusion coefficients through the entire microfilters (e) and GO membranes (f) in the presence of 30 wt% $D_2O$ in sources and drains, respectively. $Mg^{2+}$ ion diffusion coefficients through the channels within microfilters (g) and GO membranes (h) in the presence of 30 wt% $D_2O$ in sources and drains, respectively.